# LOFT - a Large Observatory For x-ray Timing

M. Feroci*[,a,b], L. Stella[c], A. Vacchi[d], C. Labanti[e], M. Rapisarda[f,a,b], P. Attinà[g],
T. Belloni[h], R. Campana[a,b], S. Campana[h], E. Costa[a], E. Del Monte[a,b], I. Donnarumma[a],
Y. Evangelista[a,b], G.L. Israel[c], F. Muleri[a], P. Porta[g], A. Rashevsky[d], G. Zampa[d], N. Zampa[d],
G. Baldazzi[i], G. Bertuccio[j], V. Bonvicini[d], E. Bozzo[k], L. Burderi[l], A. Corongiu[m], S. Covino[h],
S. Dall'Osso[c], D. De Martino[n], S. Di Cosimo[a], G. Di Persio[a], T. Di Salvo[o], F. Fuschino[e], M. Grassi[p],
F. Lazzarotto[a], P. Malcovati[p], M. Marisaldi[e], M. Mastropietro[a], S. Mereghetti[q], E. Morelli[e],
M. Orio[r], A. Pellizzoni[m], L. Pacciani[a,b], A. Papitto[l,m], L. Picolli[p], A. Possenti[m],
A. Rubini[a,b], P. Soffitta[a], R. Turolla[s], L. Zampieri[t]

[a]INAF/IASF Roma, via Fosso del Cavaliere 100, I-00133, Roma, Italy
[b]INFN Roma 2, Via della Ricerca Scientifica 1, I-00133, Roma, Italy
[c]INAF/OAR, Via Frascati 33, I-00040, Monte Porzio Catone (Roma), Italy
[d]INFN Trieste, Padriciano 99, I-34149, Trieste, Italy
[e]INAF/IASF Bologna, Via Gobetti 101, I-40129, Bologna, Italy
[f]ENEA Frascati, Via Enrico Fermi 45, I-00044, Frascati (Roma), Italy
[g]Thales Alenia Space - Italia, Strada Antica di Collegno 253, I-10146, Torino, Italy
[h]INAF/OAB, Via E. Bianchi 46, I-23807, Merate (Lecco), Italy
[i]Università di Bologna, Dip. di Fisica, Viale Berti Pichat 6, I-40127, Bologna, Italy
[j]Politecnico di Milano, Dip. di Ingegneria Elettrica, Via Ponzio 34, I-20133, Milano, Italy
[k]ISDC, Chemin d'Ecogia 16, CH-1290, Versoix, Switzerland
[l]Università di Cagliari, Dip. di Fisica, SP Monserrato-Sestu, I-09042, Monserrato (CA), Italy
[m]INAF/OACa, Loc. Poggio dei Pini, Strada 54, I-09012 Capoterra (Cagliari), Italy
[n]INAF/OACNa, Via Moiarello 16, I-80131, Napoli, Italy
[o]Università di Palermo, Dip. di Fisica, Via Archirafi 36, I-90123, Palermo, Italy
[p]Università di Pavia, Dip. di Ingegneria Elettrica, Via Ferrata 1, I-27100, Pavia, Italy
[q]INAF/IASF Milano, Via Bassini 15, I-20133, Milano, Italy
[r]INAF/OATo, Strada Osservatorio 20, I-10025, Pino Torinese (Torino), Italy
[s]Università di Padova, Dip. di Fisica, Via Marzolo 8, I-35131 Padova, Italy
[t]INAF/OAPd, Vicolo dell'Osservatorio 5, I-35122, Padova, Italy

## ABSTRACT

The high time resolution observations of the X-ray sky hold the key to a number of diagnostics of fundamental physics, some of which are unaccessible to other types of investigations, such as those based on imaging and spectroscopy. Revealing strong gravitational field effects, measuring the mass and spin of black holes and the equation of state of ultradense matter are among the goals of such observations. At present prospects for future, non-focused X-ray timing experiments following the exciting age of RXTE/PCA are uncertain. Technological limitations are unavoidably faced in the conception and development of experiments with effective area of several square meters, as needed in order to meet the scientific requirements. We are developing large-area monolithic Silicon Drift Detectors offering high time and energy resolution at room temperature, which require modest resources and operation complexity (e.g., read-out) per unit area. Based on the properties of the detector and read-out electronics that we measured in the lab, we developed a realistic concept for a very large effective area mission devoted to X-ray timing in the 2-30 keV energy range. We show that effective areas in the range of 10-15 square meters are within reach, by using a conventional spacecraft platform and launcher of the small-medium class.

**Keywords:** X-rays, high energy astrophysics, timing, compact sources, silicon drift chambers

# 1. INTRODUCTION

Neutron stars and stellar mass black holes are the densest gravitationally bound objects and possess the most intense gravitational fields in the universe. They provide a unique opportunity to investigate the properties of gravity in the strong-field regime, to reveal a variety of general relativistic effects and measure some of the fundamental properties of collapsed objects such as their radius, mass, angular momentum (see e.g. Ref.[1]). Important insights can also be gained on the properties of plasmas under extreme conditions, such as supercritical magnetic fields and bulk matter densities exceeding those of atomic nuclei. These studies will be made possible through the exploitation of several diagnostics, some of which have been singled out only in the last decade. Especially promising appear to be those diagnostics that are capable of probing *in situ* matter motion in close vicinity of collapsed objects and/or the rotation of the collapsed object itself. For instance accretion disks around neutron stars and black holes can extend down to the radius of the marginally stable orbit ($r_{ms} = 6GM/c^2$ for a non-rotating object), where velocities are comparable to the speed of light and characteristic timescales are of the order of milliseconds or less.

Besides techniques based on X-ray spectroscopy, notably very broad profiles of X-ray emission lines (Fe K-shell in particular) and the optically thick emission from accretion disks, diagnostics based on fast X-ray timing measurements hold a great potential in this respect. For example the fast quasi periodic oscillations, QPOs, that are present in the flux of a number of X-ray binary systems, most likely reflect the fundamental frequencies of the motion of matter in the innermost disk regions and thus hold the potential to gather unprecedented information on the strongly curved space-time near neutron stars and stellar mass black holes[2]. QPOs have also been observed in the X-ray flux of the very massive black holes that are hosted in Active Galactic Nuclei, AGN[3].

Direct observation of relativistic effects, such as frame dragging and strong field precession, and crucial measurements such as black hole mass and spin appear to be well within reach of the QPO diagnostics (e.g. Ref.[4]). More generally, timing variability studies have played a central role in high-energy astrophysics for a long time. The detection and accurate measurement of coherent pulsations from the rotation of neutron stars and white dwarfs yielded a wealth of crucial information on these compact objects, the torques that act on them, and the radiative processes responsible for their emission. Other periodic phenomena such as eclipses or dips are tools of paramount importance for inferring the characteristics of X-ray binaries, measuring the mass of the stars and inferring the characteristics and geometry of mass transfer in these systems.

X-ray bursts and flares, together with other impulsive phenomena of vastly different duration and characteristics have provided, among other things, evidence for thermonuclear flashes in the material accreted onto the surface of stars[5] or highly unstable field configurations inside neutron stars with exceptionally strong magnetic fields, the so-called Magnetars[6]. The fast transient oscillations detected during the most powerful flares of magnetar candidates most likely originate from seismic modes: this has opened a new perspective in the study of structure of neutron stars and the properties of ultradense and ultramagnetic matter through neutron star seismology.

While a number of important results were obtained with a variety of X-ray astronomy satellites (Uhuru, Ariel V, SAS-3, ANS, Einstein, EXOSAT, Ginga, BeppoSAX, just to mention the main ones), it was with the Rossi X-ray Timing Explorer, RXTE, that X-ray timing entered a golden age. While the RXTE discoveries were made possible by the area and time resolution of its proportional counters array (PCA, ~0.67 m$^2$), its All Sky Monitor (ASM) and flexible operations contributed greatly to the success of the mission.

The scientific goals outlined above require the development of a collimated experiment with an effective area above 10 m$^2$, good energy resolution for time-resolved spectral studies, and response extending well beyond 10 keV. Considering the success of the strategy of RXTE/PCA, we propose to follow-up the 15 years of glorious RossiXTE operations with a new mission characterised by an unprecedentedly large effective area. We believe that the new technologies will enable a giant leap forward with respect to the RXTE/PCA, thus yielding a great opportunity to fully exploit the discovery potential of timing diagnostics. In the following sections we describe the concept of a new generation timing experiment and its possible implementation.

# 2. THE CONCEPT

The issue with building extremely large area experiments resides primarily in the resources required in terms of mass, power and volume (besides, of course, cost). Our concept is based on an innovative design of large area Silicon drift chambers originally developed by INFN Trieste[7] for particle tracking in the Inner Tracking System (ITS) of the ALICE experiment in the Large Hadron Collider at CERN. The general concept of Silicon drift detectors (SDD) is to have the charge generated by the photoelectric absorption of an X-ray photon drifted through the Silicon bulk towards small collecting anodes. The small size of the read-out anodes offers a small capacitance (order of tens of fF) and therefore low noise, permitting to achieve good energy resolutions and low discrimination thresholds. The ability to drift the charge from its point of generation to the anodes makes the charge collection process independent of the photon detection: to first order, the detector performance is nearly independent on its active area (this is not strictly true because of the increasing leakage current). Some technological efforts in the field of drift detectors were aimed at achieve excellent performance in spectroscopy or timing on small area detectors. In the case of the ALICE detectors, the highest effort was devoted to maximize spatial resolution on large area monolithic devices. In fact, the ALICE-D4 detectors come with a monolithic active area of 53 cm$^2$ each and achieve spatial resolution as good as <30 µm in particle tracking[8,9]. In Figure 1 we show a picture of the ALICE-D4 detector, designed by INFN Trieste in collaboration with, and manufactured by, Canberra Inc.. It is worth noting that in the ALICE experiment <280 SDDs have operated by now for nearly 2 years; their total area is <1.5 m$^2$.

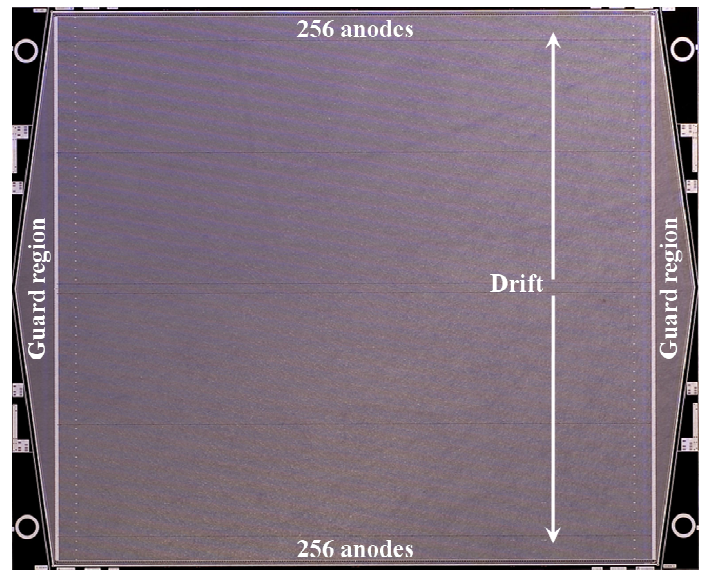

Figure 1. A picture of the ALICE-D4 silicon drift detector.

In its present design, this detector is conceptually divided into two halves (top and bottom in Figure 1), each read-out through a set of 256 anodes. A drift electric field is controlled by a voltage divider limited to a negative high voltage of <1300-2400 V. The highest voltage is applied the center of the Si tile, dropping smoothly to the grounded anodes. A side guard region is designed to gradually decrease the high voltage to the mechanical edge of the Si tile and makes approximately 17% of the total area not usable for detection purposes. The anodes have a pitch of 294 µm and a drift length of 35 mm. The read-out is then one-dimensional and each drift channel (that is, each preamplifier) subtends an area of 0.103 cm$^2$. The Si thickness (300 µm) is fully depleted.

The working principle of the detector is the following. When pairs are generated after energy release by particle ionization or photoelectric absorption or Compton scattering of photons, the electrons are first driven by the electric field to the central region of the Si bulk and then drifted through the detector until they are pulled up to the detector surface towards the collecting anodes, where a charge preamplifier transforms them into an electric signal. During the drift process, the diffusion spreads the charge cloud so that at the end of the drift its size involves more than one anode. Depending on the point of impact along the drift channel, the charge can be collected by up to 5 anodes. The charge distribution over the anodes allows the reconstruction of the photon position to a resolution up to 10 times better than the anode pitch. The maximum drift time of the charge sets the minimum time resolution. Its precise value depends on the instrument operating parameters and is about 5 microseconds.

The properties of the ALICE detectors inspired their investigation as X-ray detectors for astronomy. The first lab tests with a preliminary set-up were highly encouraging[10,11] A better, though not yet optimized, experimental set-up permitted to measure a very good performance both in terms of spectroscopy and imaging[12,13,14,15]: energy resolution between 300 eV and 600 eV (depending on the event selection) and one dimensional spatial resolution between 30 and 50 µm. The interaction point along the drift channel (that is, the second dimension) may be derived at the level of few mm (see Ref.[14] for details), even though this is irrelevant in timing applications. The low energy threshold of each read-out channel can be set to less than 500 eV, but the effective detection threshold is between 1 keV and 2 keV (depending on the accepted efficiency level), due to the charge sharing between adjacent anodes caused by the diffusion in the drift. A more detailed description of the ALICE-D4 detectors and their performance may be found in Ref.[13].

In the context of the present project, the key properties of these detectors are:

- good spectroscopic performance over monolithic large area devices
- low energy discrimination threshold
- efficient read-out (active area per unit read-out)
- low weight and volume per unit sensitive area
- proven reliable mass production
- fully scalable design
- modular approach
- full scientist control on design, production and test processes.

Given the detector performance and the INFN experience in building the 1.5 m$^2$ ALICE/ITS experiment, we began investigating the possible use and customization of this type of detectors to achieve the long-sought ten square meter experiment scale for high resolution X-ray timing in astronomy. Another major issue to be addressed is an efficient system for collimating the experiment field of view down to a typical size of <1-2º. A classical collimator would be completely unfeasible in terms of weight, volume and cost. We identify the micro-capillary technology a possible solution. Some manufacturers (e.g., Hamamatsu Photonics, Collimated Holes Inc.) produce lead-glass capillary plates. The off-the-shelf version of these devices comes with 1 mm thickness and a matrix of holes with 25 µm diameter and 32 µm pitch, providing an open area ratio of 55%. Custom design is possible (Hamamatsu Photonics and Collimated Holes Inc., private communication), reaching an open area ratio of at least 70%. These devices were proven to efficiently collimate soft X-rays both by other groups (see Ref [16]) and in our own laboratory tests[17]. We also studied the efficiency of a collimator based on this technology as a function of photon energy by means of Monte Carlo simulations. In Figure 2 we show the simulated angular and energy response. As expected the angular response depends on energy. The simulation shows that the capillary plates efficiently collimate X-rays up to 30 keV and to a wider opening angle up to <50 keV. The average density of the plates is about 2.5 g cm$^{-3}$, implying that for a 1 mm thickness their mass is 2.5 kg/m$^2$. Taking into account the current (but improvable) active/geometrical area ratio of the ALICE-D4 detectors (83%) and an open area ratio of 70%, the active to geometrical area ratio of a collimated SDD camera is 0.58.

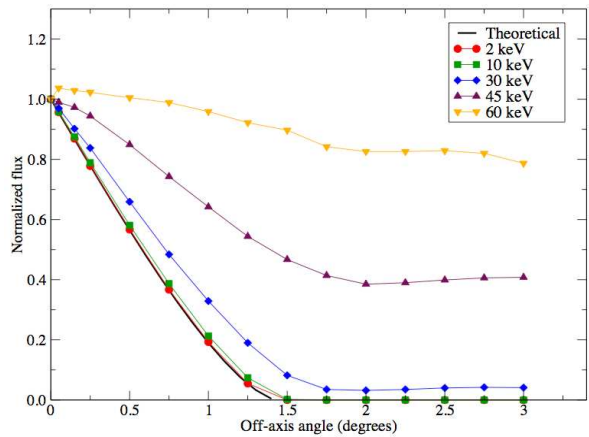

Figure 2. Angular response as a function of energy of a collimator based on a 1 mm thick polycapillary plate.

Based on what reported above, the basic element of a large area, collimated experiment is sketched in Figure 3. The sandwich is composed, from top to bottom, by a 1 mm thick capillary plate, a <500 µm thick silicon drift detector, an ASIC-based read-out and power filtering electronics board, supported by a mechanical structure, which also supports the power and data distribution. The "surface density" of such a package is of about 6 kg m$^{-2}$. This is then the basic element of a modular design for a large area experiment. Building a ten-square-meter scale experiment requires also the ability to set-up a payload of such a large area, in terms of geometrical surface, mass, power and telemetry budget.

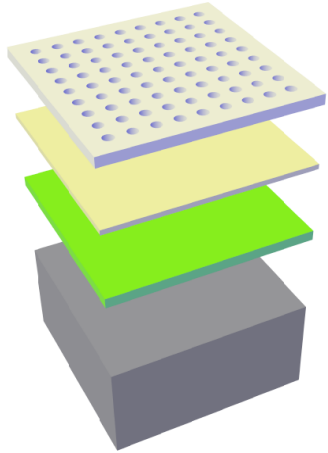

Figure 3. The conceptual basic "package" that composes the LOFT detection plane. From top to bottom: the 1 mm thick micro-capillary collimator, the 0.5 mm Si drift detector, the few mm thick front-end electronics and the supporting structure.

## 3. A POSSIBLE IMPLEMENTATION

The approach is to draw a first-order-feasible design based on conventional technology and standard bus and launchers, in a small-to-medium class satellite for a low-Earth equatorial orbit mission. We studied with Thales Alenia Space - Italia a configuration based on the PRIMA-S bus and the Vega-class launcher[18]. The first issue we studied is the maximum geometrical surface that can be deployed in orbit. Despite the thinness of the collimator-detector package, we conservatively assumed 20 cm thick detector panels, including the package itself (collimator + SDD + electronics), power distribution, thermal shielding and radiators, as well as mechanical support structure. To start with an existing and reliable deployment mechanism, we based our system on a technology used to deploy solar panels. In Figure 4 we show the satellite concept both in the deployed configuration, as well as stowed at launch inside a Vega rocket fairing. The mechanical configuration allows to host up to 32 m$^2$ of detectors geometric area. The solar panel array has a surface of 13.2 m$^2$. The detector panels can be rotated by 360º around their longitudinal axis. The solar panel array is fixed. The requirement for the alignment of the detector panels is a few arc-minutes.

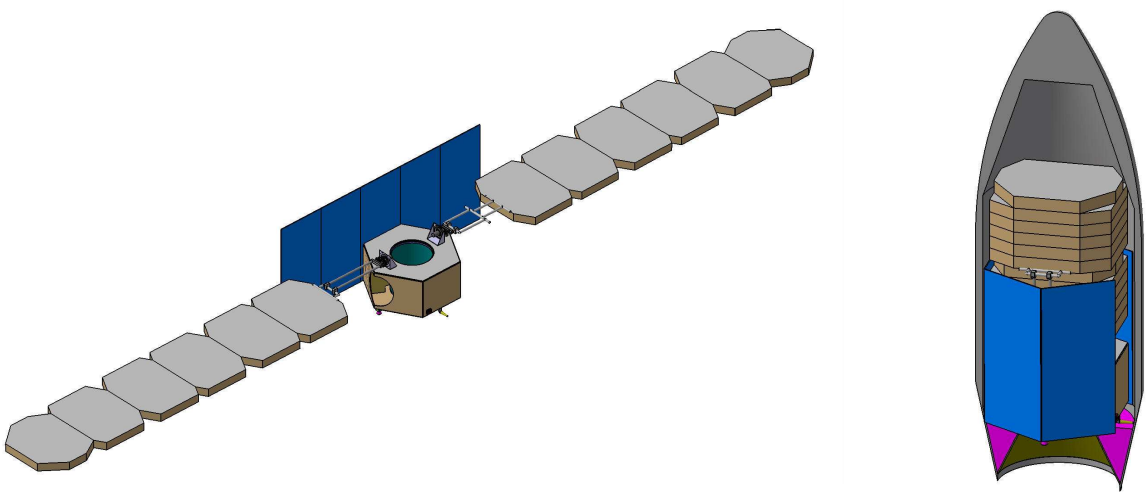

Figure 4. A possible configuration for the LOFT satellite based on a PRIMA-S bus and a Vega launcher.

We now evaluate the specific power, mass and telemetry budgets of the detector array to determine what is the active surface that can be afforded in a realistic mission. The tightest budget is given by the power. The maximum power that can be generated by the 13.2 m$^2$ solar panel array is evaluated as <4000 W. Being a low-Earth orbit satellite, we consider that half of the generated power should be reserved to recharging the batteries that power the payload when the satellite is in the Earth shadow. The power for the spacecraft services is estimated as <300 W. Assuming an 85% conversion efficiency to provide power to the payload, we end up with 1450 W available to the scientific payload.

The scientific requirements of the mission imply that in addition to the primary Large Area Detector (LAD) experiment the instrumentation includes an all sky monitor (ASM). The ASM is based on a general design[15] using the same Silicon drift detectors foreseen here for the large area experiment. A possible allocation of the ASM is shown in Figure 5; alternative options are being studied. With this configuration, the ASM is composed of 4 coded mask detector units covering a total field of view of <6-7 sr (<2 sr fully coded, <5 sr partially coded, at zero response) with an angular resolution of <7 arcmin and a point source location accuracy better than <1 arcmin. The expected sensitivity of such monitor (for each of the 4 single modules) is <0.8 Crab/s or <4 mCrab/50 ks. The total ASM mass and power budget is about 30 kg and 18 W, respectively. The power budget available to the LAD is thus approximately 1430 W.

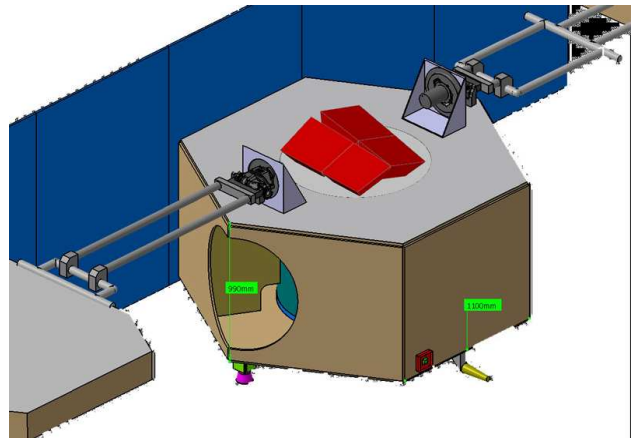

Figure 5. A possible allocation of an All Sky Monitor experiment for the LOFT mission.

The overall power budget of the LAD defines the maximum detector surface that can be simultaneously powered. In our application of SDDs to X-ray timing there is no requirement for imaging capability. The segmentation of the detector in drift channels is thus of extreme importance, but only to avoid dead-time and pile-up problems. In fact, in principle each group of 4-5 drift channels (the maximum number of anodes involved by a single event) can be taken as an independent detector. If we consider the ALICE-D4 configuration (294 µm anode pitch, 35 mm drift length), a single group of anodes covers a geometric area of the order of 0.5 cm$^2$. The drift time for each event is ≤10 µs, thus allowing a nominal maximum rate of <2x10$^5$ counts cm$^{-2}$ s$^{-1}$. Since the charge diffusion is independent of the anode pitch, this number stays roughly the same also if the pitch is made larger, doubled or more. Indeed, a larger drift channel makes the charge per channel higher (the total charge spreads over a smaller number of anodes), thus improving the low energy response (although a larger channel implies a larger dark current and a higher noise). However, assuming a detector design with 600 µm anode pitch and 35 mm drift length, the overall power per channel (including the share of the common data processing) is about 1.3 mW/channel (the ASIC alone, for which we have developed a custom design and production, requires about 0.4 mW/channel), corresponding to 6.3 mW/cm$^2$. The total 1430 W are then able to power about 22.7 m$^2$ of geometrical Silicon detector area, in turn corresponding to 13.1 m$^2$ active area. In the following we will then assume for LOFT a geometric area of 23 m$^2$ and an exposed area (accounting for the dead regions on the detector and the open area ratio of the collimator) of 13 m$^2$. We note that studies are underway to improve the effective area, both in terms of increasing the open area ratio in the capillary plates, as well as studying alternative orbits that ensure a longer exposure to the Sun, such as Sun-synchronous orbits.

We built a preliminary response matrix for the Silicon detector. We estimate an expected background rate (diffuse X-ray background plus particle-induced background) of about 2400 counts/s. The Crab Nebula is expected to give rise to 230000 cts/s. If we assume the conservative case of a smart pointing (that is pointing the satellite towards an unocculted source when a primary target goes behind the Earth) with a Crab-equivalent source always in the field of view, the count rate amounts to a telemetry budget of order 15 Mbits/s, averaged over a <6000 s orbit. This includes <2 Mbits/s from the ASM. For a low-Earth equatorial orbit and a single equatorial ground station, the satellite will be visible from the ground station for about 600 s every orbit; this translates into a downlink data-rate of <150-200 Mbit/s, well beyond the capability of S-band transponders and receivers, but within reach of X-band transponders.

We estimated the preliminary overall mass budget by making reasonable assumptions on the satellite subsystems and the payload structure. We assumed a 10% mass margin on the engineering items that already exist and have been used in previous missions. For items whose components exist but have not been integrated yet, such as the detector and electronics, the margin was elevated to 20%. A further 20% margin was then added at system level. The overall mass budget after these assumptions is 1280 kg, well within the maximum weight allocation for a Vega launcher in LEO.

Table 1 summarizes the preliminary main characteristics of the LOFT mission, as currently understood. Improvements and refinements in the numbers are expected as the design progresses.

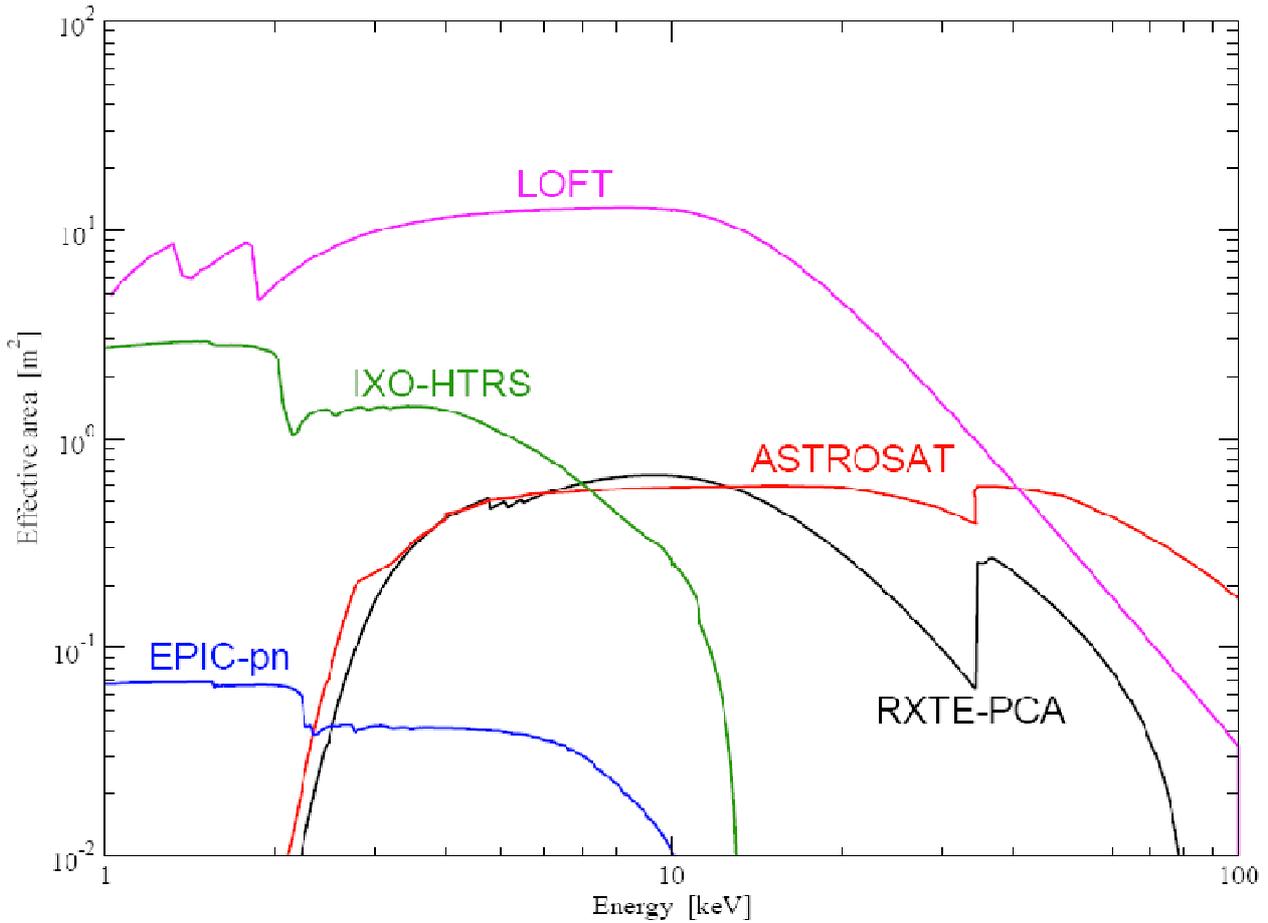

Figure 6. Estimated effective area of LOFT mission, as compared to previous (XMM/EPIC-pn, RXTE/PCA) and future missions (ASTROSAT, IXO/HTRS).

Table 1. The technical characteristics of the LOFT mission concept

| Large Area Detector - LAD | |
|---|---|
| Energy Range | 2-30 keV |
| Field of View | <1° FWHM |
| Detector Type | Silicon Drift Detectors, 450 μm thick |
| Geometric Detector Area | 23 m$^2$ |
| Effective Area | 13 m$^2$ |
| Effective Area per Channel | 0.2 cm$^2$ |
| Energy Resolution | <300-500 eV FWHM |
| Time Resolution | <10 μs |
| Dead-time (% of average rate) | <0.02% @ 1 Crab |
| Temporary Rate capability | >1000 Crab |
| Sensitivity | 50 μCrab in 50 ks - 10 mCrab in 1s |
| Mass Budget | 580 kg |
| Scientific Data Rate | 15 Mbit/s, Crab-flux obs, orbit average, smart pointing |
| Power Budget | 1400 W |
| **All Sky Monitor - ASM** | |
| Instrument type | Coded aperture Si drift detector |
| Energy Range | 2-50 keV |
| Energy Resolution | <300-500 eV |
| Geometric Area | 4 x 500 cm$^2$ |
| Field of View | >2 sr fully coded, >5 sr partially coded |
| Angular resolution | 7 arcmin |
| Point source location accuracy | <1 arcmin (S/NR>10) |
| Sensitivity (one unit, on-axis, 5σ) | 4 mCrab in 50 ks, 800 mCrab in 1s |
| Mass Budget | 30 kg |
| Scientific Data Rate | <2 Mbit/s, Crab-flux obs, orbit average, smart pointing |
| Power Budget | 20 W |
| **Overall Satellite** | |
| Payload Mass | <630 kg (20% margin included) |
| Total Mass | <1300 kg (20% system level margin added) |
| Platform | PRIMA-S |
| Launcher | Vega-class |
| Orbit | Low Earth, Equatorial |

# 4. SCIENCE WITH LOFT

The current design of the LOFT large area detector results in an effective area of <13 m$^2$, which is nearly constant between 3 and 10 keV. Below 2-3 keV, some metal and oxide implants in the present detector causes some loss of efficiency. We are studying a design that optimises the low energy response. Above <10 keV the decrease of the Si photoelectric cross section brings to a gradual decrease of the effective area; however this remains larger than 1 m$^2$ up to 30 keV. The high energy efficiency may be improved with thicker detectors, that are being studied as well. In Figure 6 we show the LOFT effective area, as compared to past and future missions. The area between 2 and 10 keV exceeds by a factor of <20 that of the largest timing experiment ever flown, the Proportional Counter Array onboard RossiXTE[19]. We now discuss how such large improvement in area impacts on the investigations of compact X-ray sources.

LOFT is mainly devoted to X-ray timing and is specifically designed to investigate collapsed objects by probing their extreme physical conditions and the space-time around them. To these aims LOFT will exploit powerful diagnostics, such as e.g. fast coherent pulsations and quasi-periodic oscillations (QPOs) in the X-ray flux of accreting neutron stars and black holes, as well as isolated magnetars. Its factor of ~20 higher effective area than the PCA on board RXTE, will prove crucial in this respect. The whole range of variability behaviour from, e.g., QPO or transient sources often requires observing times as long as weeks, i.e. the timescales over which these sources manifest different states and phenomena. A mission dedicated to timing such as LOFT will offer crucial advantages in terms of the amount of time that can be devoted to timing. In contrast, missions with a range of instruments, designed to study many classes of sources and a variety of astrophysical problems, such as the International X-ray Observatory (IXO[20]), will carry out timing studies for a small fraction of the time. LOFT and IXO are complementary also in other respects: a collimated instrument is ideal to maximise the effective area at moderate costs, and thus greatly increase the throughput of bright sources (for which spatial resolution and low background is not required). Moreover very bright events, such as gamma ray bursts and flares from magnetars, that do not occur in the satellite pointing direction and nevertheless shine through the body of the satellite and the collimator, will be recorded by LOFT with unprecedented S/N, by virtue of the very large physical area of the detector. On the other hand IXO, with its sophisticated telescope and effective area will permit to carry out unprecedented variability studies of weak sources, for which a low background is essential.

The wide-field instrument on board LOFT will monitor the X-ray sky, detect and locate bright X-ray transients, and identify source states to be followed up with pointed LOFT observations. The simultaneous availability of an all sky monitor, will be essential in this type of studies in order, e.g., to catch QPO sources in the right emission state.

Among the main subjects that LOFT will address are:

*A. The physics of collapsed objects and strong gravitational fields*

The various modes of fast quasi periodic oscillations (frequencies in the Hz to ~1 kHz range), that are often present simultaneously in the X-ray flux of accreting neutron stars and black holes, are interpreted in terms of the fundamental frequencies of particle motion in the close vicinity of the collapsed object. Even though different models make different predictions with respect to the frequencies involved, methods have been devised to test the models through detailed high signal to noise observations. For instance, in the parametric epicyclic resonance model two of the QPO modes are associated with the relativistic radial and vertical epicyclic frequencies; in the relativistic precession model instead, two of the QPO modes arise from relativistic nodal and periastron precession, while the highest frequency QPO is identified with the azimuthal frequency of motion. High precision measurements of the QPO phenomenon, such as those afforded by LOFT, will remove the degeneracy in the interpretation of the QPO modes and make it possible to exploit the corresponding relativistic frequencies in probing the space-time at a few Schwarzschild radii from the collapsed object, where strong-field effects are largest. This will permit to single out yet untested general relativistic effects, such as e.g. frame dragging, Lense-Thirring and strong-field periastron precession, verify the existence of the innermost stable circular orbit, and study gravitational light deflection in the strong field regime. It is worth emphasising that the magnitude of these effects is large and easily accessible to precise measurements once the interpretation of the QPOs is clarified. To this aim we show in Figure 7 the comparison between the 50 ks RossiXTE observation of the black hole candidate XTE J1550-564 and a 1 ks simulation in the same conditions for the source flux (<1 Crab). The "real" power spectrum displays a double peak at the frequency of 188 and 268 Hz, with fractional rms of 2.8% and 6.2% and a detection significance of 3.5 ∫ and 7.8 ∫, respectively. The improvement provided by a LOFT observation is striking. The quality of the results achieved with just 1 ks of observation demonstrates that the same phenomenon can be detected at much lower fractional rms as well as on much dimmer sources. Indeed, the whole RXTE mission has provided only a handful of detections of weak, sometimes broad, high frequency QPO signals in black hole candidates. The current indication is that the frequencies are rather stable for each system and anti-correlate with the black-hole mass. In a few cases where two peaks are observed, they appear to be at special ratios of 2:3 (and 3:5). These features have so far been detected rarely and only in a specific state when sources are bright, although this could result from a selection effect. It is not clear whether these signals and the kHz QPO in neutron star low mass X-ray binaries are the same phenomenon, as QPOs are much weaker for black holes. The observation potential of LOFT will then enable us to explore this intriguing field, for which RossiXTE could discover only the tip of the iceberg. To show this, we performed a simulation again on XTE J1550-564, where the source was artificially "evolved" to lower fluxes, while the QPOs were made weaker. The evolution was simulated according to the basic predictions of the Epicyclic Resonance Model[21] and the Relativistic Precession Model[22]. In Figure 8 we report the results of such simulations, with the source flux and fractional rms of the QPOs decreased to be as low as 300 mCrab and 0.28%, respectively.

We note that the fundamental frequencies of motion (and thus the QPO frequencies) depend on the mass and spin of the compact object and the radius at which the signals are produced. Therefore they carry also information about crucial parameters of the compact object. This can be complemented with (or checked against) estimates of the black hole mass and spin obtained by other means (e.g. the mass determination from optical spectro-photometry or the black hole spin as inferred from thermal X-ray continuum emission).

Another important diagnostics will be provided by the observation of kHz QPO oscillations in the time domain, over their coherence time scale. In order to reach the necessary sensitivity level, a large collecting area is simply essential. LOFT will afford to sample the oscillations in the time domain for a number of bright X-ray binaries and determine the shape and recurrence times of the QPO "trains". The energy resolution of LOFT, about 300-500 eV over the entire energy range, will add a new dimension to these studies. For instance by studying the energy dependence and delays in the signals it will be possible to confirm the predictions of models for the generation of QPOs by blobs orbiting in an accretion disk, measure the disk inclination and investigate strong-field lensing effects. Moreover, the resolution of 300 eV, which can be obtained by event selection, will permit to exploit also the very broad Fe-line profile and its variability as an additional diagnostic of the innermost disk regions of accreting neutron star and black holes in X-ray binaries and bright AGNs.

*B. Neutron stars structure and EOS of ultradense matter*

Understanding the properties of matter at nuclear densities and determining its equation of state (EOS) has been one of the most challenging problems in contemporary physics. Neutron stars present the best opportunity to address it. Very soft EOSs give a maximum neutron star mass in the 1.4-1.5 solar mass range, whereas stiff EOSs can reach up to 2.4 - 2.5 solar masses. Except for the case of very low masses, the neutron star radius is a powerful proxy of the EOS. Out of the different tools that have been devised to measure neutron star radii (or radius to mass ratios), there are some that rely upon X-ray spectral information, such as the gravitational redshift of lines from the neutron star atmosphere[23], the continuum spectrum and flux at the end of the contraction phase of type I X-ray bursts displaying photospheric expansion[24]. Other tools are based primarily on accurate high-time resolution measurements. Among these are: *a*. The modelling of the shape, energy dependence and delays of the pulsation from accreting millisecond spinning X-ray pulsars; owing to the fast rotational velocities and strong gravitational fields of these systems, their coherent pulsations are affected by relativistic beaming, time dilation, red/blue-shifts, light bending and, to a lesser extent, also frame dragging[25] (they provide, besides the neutron star radius and mass, an alternative means of measuring strong field gravity effects). *b*. The frequency and waveform of the fastest kHz QPOs provide also constraints on the neutron star EOS. LOFT will give great contributions to these studies through its unique combination of very large effective area and energy resolution.

A new and totally different approach to the study of the neutron star interiors and EOS has recently emerged from the discovery of global seismic oscillations (GSOs) in magnetars during two extremely luminous ``Giant Flares'' (GFs[26,27]) emitted by these sources. GSOs with frequencies ranging from tens of Hz to kHz were detected during the minutes-long decaying tail of these giant flares, when the source luminosity was about $10^{41-42}$ erg/s. The lower frequency GSOs likely arise from torsional shear oscillations of the crust[28,29,30,31]. By using them in combination with the magnetic field inferred for these magnetars it has been possible to rule out both very hard and very soft equations of state (as well as the EOS for strange stars). The possible identification of a $n = 1$ radial overtone might enable the estimate of the crust thickness. This is but an example of the diagnostic potential of GSOs. The problem with studying neutron star GSOs during giant flares is that these events are exceedingly rare (only 3 events were detected in 30 years!). While one should not miss these rare events, if GSOs could be detected during the so-called intermediate flares, i.e. shorter (1 s - 1 min long) and less energetic events from magnetars that are much more frequent than giant flares, a new perspective will emerge in the study of the neutron star structure and EOS through seismology. With its very large geometrical area LOFT is very well suited for these studies. In fact for the typical fluxes expected from intermediate flares, by observing one of such events <30° off-axis (thus, out of the collimated field of view) LOFT will collect more than $2 \times 10^5$ counts/s which can be used to look for (quasi-) periodic signals down to a pulsed fraction of better than 0.5% within 5-10s (typical pulsed fractions observed for GSOs during GFs are in the 5% - 20% range).

LOFT will be a powerful tool for studying also the X-ray variability of a very wide range of objects, from accreting and isolated pulsars, to magnetar candidates (Anomalous X-ray Pulsars and Soft Gamma Repeaters), cataclysmic variables, bright AGNs, flares stars, X-ray transients and the prompt emission of Gamma Ray Bursts. Through these studies it will be possible to address a variety of problems in the physics of high energy cosmic sources. In combination with multiwalength campaigns LOFT can also address the interplay of accretion and jet ejection, a still poorly understood phenomenon occurring over vastly different conditions, ranging from pre-main sequence stars to the massive black holes of AGNs.

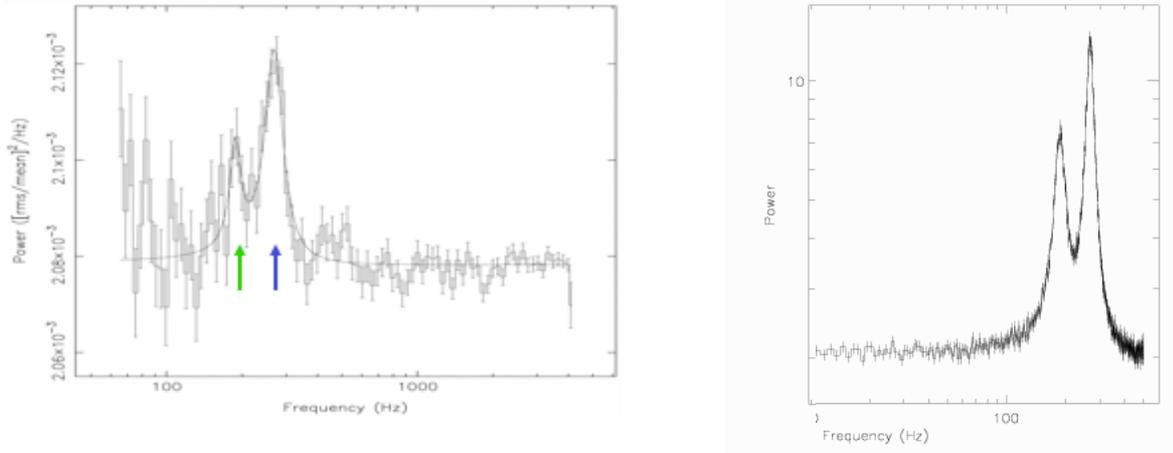

Figure 7. *Left:* The high frequency QPOs detected by RXTE/PCA in the black hole candidate XTE J1550-564.[32] The 50 ks PCA observation led to a 7.8 and 3.5 ∫ detection of the 188 and 268 Hz peaks, respectively. *Right:* the same source in the same conditions in a simulated observation with LOFT, with an exposure of only 1 ks.

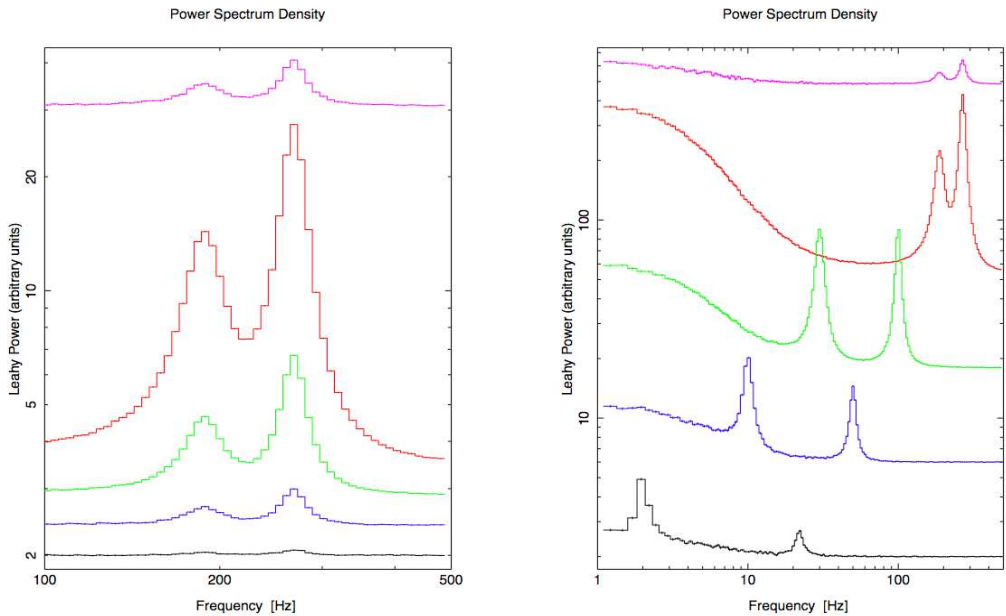

Figure 8. Simulated 30 ks LOFT observations of XTE J1550-564, with the source artificially varied in intensity and fractional rms of the QPOs. The source evolution was simulated according to the predictions of the Epicyclic Resonance Model (*left*) and the Relativistic Precession Model (*right*), to demonstrate the potential of LOFT in understanding the origin of these phenomena. Power spectra are vertically shifted for the sake of clarity. The topmost curve is a RXTE simulation. The other curves correspond to (top to bottom): flux 1 Crab, fractional rms 2.8% and 6.2% (as in Figure 7); 800 mCrab, 1.4% and 3.1%; 600 mCrab 0.7% and 1.5%; 300 mCrab, 0.28% and 0.62%.

## 5. CONCLUSIONS AND PERSPECTIVES

The Silicon Drift Detectors developed at INFN Trieste and the micro-capillary plates as collimators appear to be powerful enabling technologies for the development of an extremely large area experiment. In the current design we reached an effective area of 13 $m^2$, in the context of a small-medium satellite, based on a PRIMA-Science platform for a low Earth orbit. Such configuration is currently technically limited by the power budget. This is expected to improve in terms of optimization of the power consumption of the proposed device. Moreover the geometric/exposed area ratio, as determined by the inactive detector areas and the covering factor of the micro-capillary plates, is currently 58% and can probably be improved. We will work on both issues to increase the instrument effective area. Finally, an orbit with a higher exposure to the Sun, with respect to the 50% offered by a LEO would also lead to a higher effective area.

Future perspectives for X-ray timing are unclear at present. RossiXTE is close to end of operation and the only planned mission that offers an area that is sufficient for timing studies is the Indian satellite ASTROSAT[33], which presents with a somewhat larger area than RossiXTE only at energies > 15 keV . The need for a major step forward is clear to the entire community. The International X-ray Observatory (IXO) is expected to carry the High Time Resolution Spectrometer (HTRS[34]). However, the HTRS instrument will only be able to operate for a small fraction of time, being alternative to the main experiments onboard the same observatory. Moreover, IXO's ~ 2 $m^2$ effective area at 1-2 keV decreases significantly at higher energy, reaching <0.6 and 0.3 $m^2$ at 6 and 10 keV, respectively (see Figure 6), even though the low background due to the imaging telescope will be a significant advantage for weak sources. LOFT will be an extremely powerful satellite to study relatively bright sources (i.e. brighter than a few mCrab); for these sources collecting photons with a very large effective area and carrying out long observations to follow to evolution of the properties across different states are among the most important goals. In addition, by its concept, LOFT will be able to respond rapidly to interesting events or states of variety of sources, through its own monitoring of the sky, and will be able to study the timing properties in a wide energy range, from 2 to 30 keV. Its large geometric area will provide a very high throughput also for hard and very bright events occurring outside its field of view, such as flares from magnetars or gamma ray bursts. Based on the technology readiness, LOFT can be built for a flight in the second half of this decade.